\documentclass[useAMS,usenatbib]{mn2e}
\usepackage{graphicx}
\usepackage{textcomp}
\usepackage{amssymb}\usepackage{hyperref}

 \usepackage{times}

\newcommand\lsim{\mathrel{\rlap{\lower4pt\hbox{\hskip1pt$\sim$}}
        \raise1pt\hbox{$<$}}}
\newcommand\gsim{\mathrel{\rlap{\lower4pt\hbox{\hskip1pt$\sim$}}
        \raise1pt\hbox{$>$}}}
\newcommand{\be}{\begin{equation}}
\newcommand{\ba}{\begin{eqnarray}}
\newcommand{\ee}{\end{equation}}
\newcommand{\ea}{\end{eqnarray}}

\newcommand{\Msun}{\,{\rm M_\odot}}
\newcommand{\yr}{\,{\rm yr}}

\title[Stars as resonant absorbers of gravitational waves]{Stars as resonant absorbers of gravitational waves}

\author[B. McKernan, K.E.S.Ford, B.Kocsis, Z.Haiman]{B. McKernan$^{1,2,3,4}$\thanks{E-mail:bmckernan at amnh.org (BMcK)}, K.E.S. Ford$^{1,2,3,4}$, B.Kocsis$^{5,6}$,Z.Haiman$^{7}$\\
$^{1}$Department of Science, Borough of Manhattan Community College, City University of New York, New York, NY 10007, USA\\
$^{2}$Department of Astrophysics, American Museum of Natural History, New York, NY 10024, USA\\
$^{3}$Graduate Center, City University of New York, 365 5th Avenue, New York, NY 10016, USA\\
$^{4}$Kavli Institute for Theoretical Physics, UC Santa Barbara, CA 93106, USA\\
$^{5}$Harvard-Smithsonian Center for Astrophysics, 60 Garden St., Cambridge, MA 02138, USA\\
$^{6}$Institute for Advanced Study, Einstein Drive, Princeton, NJ 08540\\
$^{7}$Department of Astronomy, Columbia University, 550 West 120th Street, New York, NY 10027\\}

\begin{document}

\date{Accepted. Received; in original form}

\pagerange{\pageref{firstpage}--\pageref{lastpage}} \pubyear{2008}

\maketitle

\label{firstpage}

\begin{abstract}
Quadrupole oscillation modes in stars can resonate with incident
gravitational waves (GWs), and grow non-linear at the expense of GW
energy.  Stars near massive black hole binaries (MBHB) can act as
GW-charged batteries, discharging radiatively. Mass-loss from these stars
can prompt MBHB accretion at near-Eddington rates. GW opacity is
independent of amplitude, so distant resonating stars can eclipse GW
sources. Absorption by the Sun of GWs from Galactic white dwarf
binaries may be detectable with second-generation space-based GW
detectors as a shadow within a complex diffraction pattern.
\end{abstract}


\section{Introduction}
Supermassive black holes (SMBH) with masses in the range $\sim 10^{6}-10^{9}M_{\odot}$ 
are present in the nuclei of most, perhaps all, nearby galaxies \citep[see e.g. the recent 
review by][]{KormHo13}. Mergers between galaxies should result in supermassive black 
hole binaries; indeed active SMBH binaries have been directly resolved at 0.1-1kpc 
separations in X-rays \citep{Kom03,Fab11}, from sub-kpc to a few kpc separations in 
the optical band \citep{Woo14,Comerford13,Comerford14}, and at $\sim 10$pc separation 
in the radio \citep{Rod06}. Most of the binding energy of a merging massive binary is radiated as
gravitational waves \citep{ThBrag76}. As the binary approaches merger, the gravitational wave (GW) 
frequency ($\nu_{\rm GW}$) increases in a chirp, passing through quadrupolar
($\ell=2$) oscillation frequencies ($\nu_{\ast}$) of stars and stellar remnants, 
resonating whenever $\nu_{\rm GW} \sim \nu_{\ast}$. The interaction of GWs with matter has been considered in various contexts, 
\citep[e.g.][]{Hawk66,KocsisLoeb08,Li12}; the latter suggesting that viscous heating of Sun-like stars by GW from a 
nearby merging massive black hole binary can reach $\sim L_{\odot}$. However, resonant interactions of 
GWs with normal, as opposed to compact stars (similar to a bar detector), has not received very much attention \citep{MTW,Chandra91,Chandra92,siegelroth10,siegelroth11}. It has been shown that GW can do work on stellar oscillations 
leading to potential observable effects on the oscillations \citep[e.g.][]{FabGough84,Kojima05}.
After this manuscript was submitted, a pre-print appeared on arXiv.org by \citep{LopesSilk14}, considering the resonant interaction of GWs with stars, as well as assessing the feasibility of detecting the induced stellar oscillations through astroseismological measurements. 
In this Letter, we discuss the possibility of GW absorption lines at resonant frequencies in stars, 
eclipses of GW sources by foreground stars (including the Sun) and the possible use of stars
 in galactic nuclei as electromagnetic detectors of resonating GW from nearby massive black hole binaries. In the latter case, we show that \emph{resonant} heating of a single mode in a Sun-like star can be up to $\sim$11 orders of magnitude larger than the viscous heating in \citep{Li12}. 

\section{GWs from a binary resonating with stellar oscillations.}  A
circularized binary with individual BH masses $M_1$ and $M_2$ and
physical separation $a_{\rm bin}$ emits GWs at frequency
\begin{equation}
\nu_{\rm GW}=\frac{2}{t_{\rm orb}}=\frac {G^{1/2}M_{\rm bin}^{1/2}}{\pi a_{\rm bin}^{3/2}} =
 2 \, M_6^{-1} a_{1}^{-3/2}\, {\rm 1 mHz},
\label{om_gw}
\end{equation}
for a characteristic duration
\begin{equation}
t_{GW}=\frac{a_{\rm bin}}{|\dot{a}_{\rm bin}|} =
0.8\, \eta_{-3}^{-1} M_6^{-5/3} \nu_{\rm GW,1}^{-8/3} \, \yr,
\label{decay}
\end{equation}
where
$t_{\rm orb}$ is the orbital period,
$M_{\rm bin}=M_{1}+M_{2}$ is total binary mass,
$r_{g}=GM_{\rm bin}/c^{2}$ is the gravitational radius of the binary,
$a_1\equiv a_{\rm bin}/(10 r_{\rm g})$,
$M_6\equiv M_{\rm bin}/10^6 \Msun$,
$\eta \equiv M_1M_2/(M_1+M_2)^2$ is the symmetric mass ratio,
$\eta_{-3}\equiv \eta/10^{-3}$,
$\nu_{\rm GW,1}\equiv \nu_{\rm GW}/1{\,\rm mHz}$
and where the orbital decay $\dot{a}_{\rm bin}$ is driven by the
(quadrupolar) GW emission \citep{Peters&Mathews63}.  The resulting GW
strain amplitude averaged over directions is given by
\begin{equation}
h=\sqrt{\frac{32}{5}}\frac{G^{2}}{c^{4}} \frac{M_{\rm bin} \mu_{\rm bin}}{D_{\ast} a_{\rm bin}} =
1.6\times10^{-7} \nu_{\rm GW,1}^{2/3} M_6^{-1/3} \mu_3 D_{\ast,3}^{-1}
\label{strain}
\end{equation}
where $\mu_{\rm bin}$ is the reduced mass, $D_{\ast}$ is the
resonating star's distance from the binary, $\mu_{3}\equiv \mu_{\rm
  bin}/10^{3}\Msun$ and $D_{\ast,3}\equiv D_{\ast}/10^3 r_g$. Sun-like
stars have $\ell=2$ oscillation modes with frequencies $\omega_{\ast}
= 2\pi \nu_{\ast}$ spanning $10\mu$Hz$-0.1$Hz \citep{astseism10}
\footnote{ For the solar model in Fig. 3.20. in \citep{astseism10},
$\nu_{n} \approx 0.14\,(n+2){\,\rm mHz}$ and $1.5\,(n+3)^{-1}{\,\rm mHz}$
approximately for $p$ and $g$-modes, respectively,
for $n\leq 30$.}, which
can match the frequency of GWs from a binary source.  Tens of low-radial-order
f, g, and p-modes with overlap integrals $\gtrsim 10^{-3} M_{\ast}$ span
$\sim 0.1-1$mHz in solar models \citep{astseism10}.

We follow the approach and definitions of \citet{Rathore05} in representing GW-driven oscillations of a stellar mode by a driven
damped harmonic oscillator, whose displacement $x(t)$ is the solution
to
\begin{equation}
\ddot{x}+\frac{\dot{x}}{\tau_{d}} +\omega_{\ast}^{2} x=F(t),
\label{ddho}
\end{equation}
where $\tau_{d}$ is the damping time of the stellar mode and $F(t)=F_{\rm GW}(t)$ is
the driving force. Low-order g-modes in the linear regime damp radiatively on the
timescale, $\tau_{d} \sim 10^6\,$yr \citep{kumargoodman96}.  However,
in the nonlinear regime (with energy in the mode $E_m\gsim10^{37}$erg),
 coupling to high-degree g-modes reduces this timescale to
$\tau_{d} \approx 50 E_{42}^{-1/2}$day, (here
$E_{42}=E_m/10^{42}$erg).  For f-modes in convective stars, linear
dissipation through turbulent viscosity takes $\sim 10^{4}$yr
\citep{Rayetal1987}, but in the nonlinear regime, dissipation via high
order p-modes (of degree $\ell=0,2,4$) occurs on timescale $\tau_{d} =
3 \times 10^{4}E_{42}^{-1}$day \citep{kumargoodman96}.  Multiplying
eq.~(\ref{ddho}) by $\dot{x}$ and integrating over time yields the
familiar form, expressing conservation of energy

\begin{equation}
\dot{E}_{m} +\dot{Q}=\dot{W},
\end{equation}
where $E_{m}$ is the mechanical energy, $Q$ is the energy lost via
dissipation, and $W$ is the work done by the driving force,
\begin{equation}
 E_{m}=\frac{\dot{x}^{2}}{2}+\frac{\omega_{\ast}^{2}x^{2}}{2},\;\; Q=\int^{t}_{t0}
\frac{\dot{x}^{2}}{\tau_{d}} dt,\;\; W=\int^{t}_{t0} F(t) \dot{x} dt,
\end{equation}
respectively. All three quantities are per unit mass in a single mode.

Excitation of non-radial oscillations in stars and stellar remnants due 
to tidal capture is relatively well-studied \citep[e.g.][]{pt77,rg94,Rathore05}, 
compared to oscillation excitation due to incident GWs. In the latter case, 
the effective driving force per unit mass due to GWs is nearly
sinusoidal, with a characteristic amplitude $|F_{\rm GW}| =
\omega_{GW}^{2}|h|R_{\ast}$
\citep{MTW,KS97,siegelroth10,siegelroth11}.  
The frequency of $|F_{\rm GW}|$ during the inspiral phase of a binary
evolves slowly (i.e. the number of orbits at $\nu_{\rm GW}$ is
$N=\nu_{\rm GW}^2/\dot{\nu}_{\rm GW}\gg 1$).  In this limit, $|F_{\rm
  GW}|$ is a sinusoid of nearly constant amplitude but slowly
increasing frequency $\nu_{\rm GW,0}+\dot{\nu}_{\rm GW, 0}t$. It can
be shown that in the absence of damping
($\tau_d\rightarrow\infty$), the effective duration of the resonant
forcing, while the source drifts across a resonance, is $t_{\rm F}
\approx 1/\sqrt{4\dot{\nu}_{\rm GW}}$ \citep[e.g.][]{Rathore05}, yielding
\begin{equation}
\label{eq:tF}
t_F=6.6 \left(\frac{M_{\rm ch}}{\Msun}\right)^{-5/6}
\left(\frac{\nu_{\rm GW}}{\rm 1 mHz}\right)^{-11/6} \, \yr.
\end{equation}
Here $M_{\rm ch}=\eta^{3/5}M_{\rm bin}$ is the chirp mass.  Analytic
solutions to eq. (\ref{ddho}) can be found in two limiting cases: the
saturated/steady-state case with constant forcing frequency ($t_{\rm
  F} \gg \tau_{d}$) \citep{MTW} and the undamped case ($t_{\rm F} \ll
\tau_{d}$) \citep{Rathore05}.  Expressing the damping time of a given
stellar oscillation mode in terms of the ``quality factor'' $q_{\rm
  f}=\omega_{\ast} \tau_{d}/\pi$, the saturation condition
$t_F=\tau_d$ implies that steady-state is reached approximately for
\begin{equation}
M_{\rm ch} \lesssim 0.35 \left(\frac{\nu_{\rm GW}}{\rm
  1 mHz}\right)^{-1} \left(\frac{q_{\rm f}}{10^6}\right)^{-6/5} \,
\Msun.
\label{saturation}
\end{equation}
Steady-state will not be reached for $q_{\rm f} \gg 10^6$. However, if
the star is close to the GW source such that $|F_{\rm GW}|$ is
sufficiently large, the oscillations can grow nonlinear before
reaching the steady-state limit.  If so, the mode coupling to
higher-order modes prohibits further growth, and the effective quality
factor is greatly decreased.

\section{Saturated/Steady-state limit ($t_{\rm F} \gg \tau_{d}$).}
Assuming stationary GW forcing at a constant frequency
($F=|F|e^{i\omega_{\ast}t}$), the maximum steady-state displacement
$x_{\rm max}$ is
\begin{equation}
x_{\rm max}= \frac{|F|}{\sqrt{(\omega_{\ast}^{2}-\omega_{\rm
      GW}^{2})^{2} +(\omega_{\rm GW}/\tau_{d})^{2}}}
\label{xmax}
\end{equation}
or $x_{\rm max} \approx |F|\tau_{d}/\omega_{\ast}=\pi R_{\ast} q_{\rm
  f} h$ in the limit $(\omega_{\ast}^{2}-\omega^{2})^{2} \ll
(\omega/\tau_{d})^{2}$. In the steady-state solution, $\dot{E}_{m}=0$
and the cycle-averaged power of the external forcing $\langle \dot{W}
\rangle=\langle F\dot{x}\rangle$ equals the rate of heating $\langle
\dot{Q} \rangle=\langle \dot{x}^{2} \rangle/\tau_{d}$. Taking the
limit $\omega_{\rm GW} \approx \omega_{\ast}$, the rate of work done
in the steady-state case is $\dot{W}_{s} \approx \langle
F^{2}\rangle\tau_{d}$ or
\begin{equation}
\dot{W}_{s}=\dot{Q}_{s} \approx \frac{\pi}{2} R^{2}_{\ast}h^{2}q_{\rm f}\omega_{\ast}^{3}.
\end{equation}
The cross-section for absorbing GWs is given by
\begin{equation}
\sigma_{\rm GW}=\frac{M_{m}\langle \dot{W}_{s} \rangle}{\Phi_{\rm GW}} \approx \frac{8\pi G}{c^{3}}M_{m}R_{\ast}^{2}\omega_{\ast}^{2}\tau_{d}
\end{equation}
where $\Phi_{\rm GW}=(c^{3}/16\pi G)\dot{h}^{2}$ is GW flux incident
on the star and $M_{m}$ is the overlap with the normal mode expressed
as a measure of the mass involved in the mode such that \citep{KS97}
\begin{equation}
  M_{m} x \equiv \left(\int \xi \rho h \nabla V d^3 x\right)/ \int \rho
  |\xi|^2 d^3 x
\end{equation}
where $\xi=\xi_{nlm}(r)$ is the displacement for a normal mode and 
$V=\frac{1}{2}(x^2 - y^2)$. The fractional energy
flux removed from the incident GW, corresponds to a resonant 'optical depth' ($e^{-\tau}$). 
The ``effective opacity'' seen by the GWs in the steady-state limit is $\tau_{\rm
  eff,s}\equiv\sigma_{\rm GW}/\pi R_{\ast}^{2}$ or
\begin{equation}
\tau_{\rm eff,s} \approx \frac{8\pi G}{c^{3}}M_{m} q_{\rm f}\omega_{\ast}=0.8 \left(\frac{\nu_{\ast}}{1\rm{mHz}}\right)\left(\frac{q_{\rm f}}{10^{6}}\right)\left(\frac{M_{m}}{\Msun}\right).
\label{atten}
\end{equation}
 In general, computing overlap integrals ($M_{m}$) between stellar modes and
the GW forcing for realistic stellar structure models will be
difficult, and will also be very sensitive to the details of stellar
structure.  A full investigation is beyond the scope of this Letter.
However, overlap calculations exist for the somewhat similar case of
Newtonian tidal forcing by a nearby point-source, both for simplified
polytropes \citep[e.g.][]{pt77,rg94} and for more realistic stellar models \citep{astseism10}. 
In general, these show that the lowest-order modes
have large overlap integrals, between O(0.1)-O(1) for polytropes (see,
e.g. Table 1 in \cite{pt77}), but also that simple polytrope
models are insufficient to estimate the excitation of g-modes in
Sun-like stars \citep[e.g.][]{W12}.  \cite{KS97} calculate overlap integrals for
GW forcing of polytropes, showing that the fundamental mode has an
overlap integral between $20\%-40\%$ for simple polytropic fluid models,
with polytrope index $1.5<n<2.5$ (see their Table 1, where the fundamental 
f-mode is labeled as $p_{1}$).  We
find this result intuitiviely unsurprising, since the angular part of
the overall integral, for ell=2 modes, matches the quadrupolar pattern
of the GWs, and the radial integral is over the product of a
non-oscillatory eigenmode and a slowly-varying GW forcing function.
We conclude that the overlap integral for a number of low-radial-order
ell=2 modes is likely to be significant, i.e. close O(0.1)-O(1) in at
least a few cases, depending on the details of stellar structure.
More sophisticated stellar modelling is needed to compute the overlap
for g-modes, and also for non-solar type giant stars, where GW
wavelength is closer to the stellar radius and where higher-order
modes may have substantially greater overlap integrals. 
\citep{LopesSilk14} calculate Solar models beyond a simple polytrope and 
find that values of $M_{m}$ ($\Xi_{n}$ from their eqn. 9) for p-modes and f-modes are supressed by $1-2$ 
orders of magnitude relative to the values in 
\cite{KS97} because of the rapid change in density profile. Stellar structures closer 
to simple polytropic models (e.g. red giants or fully convective low mass stars) 
may therefore be the most efficient at GW absorption.

In the linear regime, low-order $\ell=2$ modes with $q_f\sim {\rm
  few}\times 10^{7}$, $\nu_\ast \sim{\rm few}\times 0.1$mHz, driven by
a long-lived $\sim M_{\odot}$ binary, can approach saturation and
attenuate GWs significantly. Work done by GWs
on resonant stellar oscillation modes can extract a significant
fraction of incident GW energy, far exceeding nonresonant viscous GW
dissipation \citep{Li12}. At higher $\nu$ (and $q_f$), the mode
will not saturate (eq.~{\ref{saturation}), whereas for much smaller
  $q_f$ (e.g. for stars near MBHBs, driven nonlinear), the saturated
  opacity is small.

\section{Undamped limit ($t_{\rm F} \ll \tau_{d}$).} Undamped
($\tau_{d} \rightarrow \infty$) stellar oscillations driven by a slowly
varying frequency ($\dot{\omega}\ll\omega^2$) acquire energy equal to
$E_{m}=|F|^{2}t_{F}^{2}/2$ \citep{Rathore05}.  This energy is then dissipated on the
longer timescale $\tau_d\gg t_{\rm F}$, so that $Q_u=E_m$.  The
average rate of work done during the forcing is
$\dot{W}_{u}=W_{u}/t_{F}=E_{m}/t_{F} \approx |F|^{2}t_{F}/2$. Thus
$\dot{W}_{u} \approx \dot{W}_{s}(t_{F}/\tau_{d})$, and the average
effective opacity to GWs, while the modes are being resonantly driven,
is $\tau_{\rm eff,u}=\tau_{\rm eff,s}(t_{F}/\tau_{d})$:
\begin{equation}
\tau_{\rm eff,u}=0.3 \left(\frac{M_{\rm ch}}{\Msun}\right)^{-5/6}\left(\frac{\nu_{\ast}}{1\rm{mHz}}\right)^{1/6} \left(\frac{M_{m}}{\Msun}\right),
\label{attenu2}
\end{equation}
independent of $q_{\rm f}$. The mean dissipation rate is
$\dot{Q}_{u}=\dot{Q}_{s}(t_{F}/\tau_{d})$. For large but finite
$\tau_{d}$, both $\tau_{\rm eff, u}$ and $\dot{Q}_{u}$ approach the
maximal saturated case as $t_{F} \rightarrow
\tau_{d}$. Eqn.~(\ref{attenu2}) implies that GWs from stellar-mass
binaries can be strongly attenuated by resonant low-order stellar f-modes.

\section{Eclipses of GW sources by the Sun or stars in the MBHB host galaxy.}
\label{sec:eclipses}
The opacities $\tau_{\rm eff,u(s)}$ are independent of $D_{\ast}$, and
g-, f-, and p-mode frequencies for the Sun are coincidentally in the
sensitivity band of the proposed eLISA instrument \citep{cutler98}. The
Sun could therefore annually eclipse eLISA GW sources located in the
ecliptic plane - in particular, white dwarf binaries (WDBs)
\citep{Crowder07}. The effect would be a ``shadow'' within a complex
GW diffraction pattern near the resonant frequency (since the GW
wavelength exceeds 1AU). Using the Monte Carlo simulations of \citep{Timp06}, 
we estimate (from their Fig. 12) that between 20 (SNR $>=5$) and 5(SNR$>=10$)
individually resolvable WDBs with LISA in a 1 year observation will
lie in a $3.3\mu $Hz bin (corresponding to width $d\nu/\nu~1/q_f~1/100$) 
around log(-3.5) Hz, i.e.near prominent Solar modes. Given that the plane
of the ecliptic is ~1/360th of the sky, this gives a $\sim2\%-6\%$ chance
that one such WDB will be occulted annually by a low-order p-mode of the Sun
annually. Using the 3 largest mass Solar p-modes (listed in
Table 1 in \citep{CutLind96}), each with
$q_{f} \sim 100-400$, we find odds of $\sim 1\%-6\%$ for each of the modes that a WDB
could lie in the ecliptic plane at that frequency in a $3.3\mu$Hz wide
bin. The overwhelming majority of WDBs will lie off the ecliptic, but the orbits of
future space-based GW detectors may be chosen to allow the most
promising eclipses to be observed. We estimate that eLISA will have an O(10\%) chance of identifying a WDB near a
Solar resonance with a deep O(0.1)-O(1) transit depth that a future space-based GW detector could observe.  The chances
increase significantly for higher-order modes with much broader
resonances \citep{Stix+1993} but owing to their low overlap integrals and/or
low $q_f$, these modes will likely produce much shallower transits.

GW absorption could also be detected as a result of transits by
bloated stars in the $\sim$ 35,000 galactic nuclei within
50Mpc or extreme mass-ratio inspirals around Sgr A* \citep{Amaro07}. Conservatively 
assuming a $1\%$ AGN rate, $\sim 350$ active
galaxies lie within the LISA search window for transits. From \citep{McK12} 
most of these AGN host SMBH-IMBH or SMBH-sBH (EMRI) binaries. For $\sim 10$ such binaries
in the LISA frequency window ($\sim 0.1-$few mHz) we estimate a probability of
$0.01-1$ that we would see one transit in a 10-year mission, assuming
$0.01\%-1\%$ chance of a transit/AGN/yr \citep{bk13}. EM
study of such transits would be challenging but potentially detectable \citep[e.g.][]{MY98,TM09}. 
However, the resonant driving by these systems lasts for only $t_{F}\lsim$ day (eq.~\ref{eq:tF}) so the chance of
GW absorption coinciding with EM transits will be negligible.

\begin{figure}
\vspace{-1.2\baselineskip}
 \includegraphics[width=8.0cm,clip=true,angle=0]{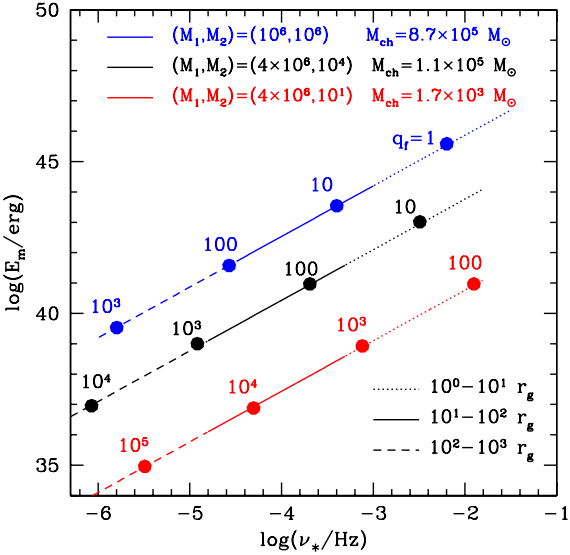}
\vspace{-1\baselineskip}
\caption{\label{fig:freq} Total energy ($E_m$) deposited in a single
  resonant mode at frequency $\nu_\ast$ of a Sun-like star, when an
  inspiraling binary GW source sweeps across this frequency. The star
  is located at $D_{\ast}=10^{3}r_{g}$ from a merging MBHB.  Three
  different BH mass combinations are shown, and we assume $M_{\rm m}
  \sim M_{\ast}$. Values of $q_{\rm f}=2\nu t_{\rm F}$ are indicated
  for saturation at the corresponding frequency.  }
\end{figure}

\section{Resonant GW-heating of stars.}  
A star orbiting near a merging MBHB (within $\sim$1 pc) can absorb a
significant amount of resonant GW energy. The
average undamped heating rate of a single mode during the passage
through resonance is $M_{m}\dot{Q}_{u}=(1/2) M_{m} |F_{GW}|^2 t_{\rm
  F}$.  As an illustrative example, we consider an $M_2=10^4 {\rm
  M_\odot}$ IMBH separated by $\approx 15 r_g$ from Sgr A*
($M_1=4\times 10^6 {\rm M_\odot}$), and a Sun-like star $10^3 r_g$
away, resonating with the GWs (at $\nu\approx 0.3$ mHz):
\begin{eqnarray}
&&M_{m}\dot{Q}_{u} = 400 L_{\odot}\;
\left(\frac{M_{m}}{\Msun}\right)
\left(\frac{R_{\ast}}{R_{\odot}}\right)^{2}
\left(\frac{D_\ast}{10^{3}r_{g}}\right)^{-2}\times\nonumber\\ &&
\left(\frac{M_{\rm bin}}{4\times10^6\Msun}\right)^{-1}
\left(\frac{\mu}{10^4\Msun}\right)^{3/2}
\left(\frac{\nu}{\rm{0.3mHz}}\right)^{7/2}.
\label{ssheat}
\end{eqnarray}
The heating rate is high, but lasts only for $t_F\sim 1.5$\,days for
the fiducial parameters for a single mode.  The corresponding energy
dumped into the mode is 
\begin{eqnarray}
&&E_m=M_{m}{Q}_{u} = 10^{41}\, {\rm erg}\;
\left(\frac{M_{m}}{\Msun}\right)
\left(\frac{R_{\ast}}{R_{\odot}}\right)^{2}
\left(\frac{D_\ast}{10^{3}r_{g}}\right)^{-2}\nonumber\\ && \times
\left(\frac{M_{\rm bin}}{4\times10^6\Msun}\right)^{-4/3}
\left(\frac{\mu}{10^4\Msun}\right)
\left(\frac{\nu}{\rm{0.3mHz}}\right)^{5/3}.
\label{eq:ssheattotal}
\end{eqnarray}
Fig.~\ref{fig:freq} shows the total energy $E_{m}$, deposited in a single resonant mode of a star for the fiducial values of $M_{m}=M_{\odot},R_{\ast}=R_{\odot},D_{\ast}=10^{3}r_{g}$ for three different MBHBs. From Fig.~\ref{fig:freq}, for a large overlap integral, up to $10^{45}$erg can be deposited into a single mode of a star near an equal mass ($10^{6}M_{\odot},10^{6}M_{\odot}$) MBHB. This is $\sim 11$ orders of magnitude larger than the expected viscous heating of stars \citep{Li12}. If this much energy can emerge on short timescales, the resonating star can act as a prompt electromagnetic signposts of incident GWs.

\section{Rate of discharge of GW-charged batteries.} 
 Stars can release
$E_m$ either electromagnetically (EM) or via GW emission (equivalent
to elastic scattering of incident GWs). The GW timescale $\approx
5c^{5}/(G E_m \omega_{GW}^{2})\gg$ EM timescales.  Energy thermalized
in radiative zones emerges on the thermal timescale $\tau_{\rm th}$
($\approx 10^7{\rm yr}$ for Sun-like stars), implying that massive
stars will brighten on this long timescale. The fractional luminosity
increase is limited to $E_{m}/E_\ast\ll 1$.  However, energy deposited
in the convection zone emerges on timescale $t_{\rm
  conv}\sim 10^6{\,\rm s}$, which may cause significant brightening
(see below).

\subsection{Resonant destruction of stars by GWs.} 
A star is completely
disrupted when the total energy dumped into the star
(eq.~\ref{eq:ssheattotal}) becomes greater than the binding energy of the
star $E_{\ast}\sim GM_{\ast}^{2}/R_{\ast}$. In the limit $M_{m} \sim
M_{\ast}$ this happens at a radius $D_{\rm rd}\gg r_g$ where
\begin{equation}
\frac{D_{\rm rd}}{D_{\rm td}} = 0.03 \left(\frac{M_{\ast}}{M_{\odot}}\right)^{-1/6} \left(\frac{R_{\ast}}{R_{\odot}}\right)^{1/2}
\left(\frac{\nu_{\ast}}{{\rm 1 mHz}}\right)^{5/6} \left(\frac{\mu}{10^4 \Msun}\right)^{1/2}
\label{disrupt}
\end{equation}
where $D_{\rm td}=R_{\ast}(M/M_{\ast})^{1/3}$ is the tidal disruption
radius. Solar type stars near MBHBs are thus disrupted by Newtonian tides well
before destruction due to resonant GW absorption.  

\subsection{Near-field destructive effects.}
Far from MBHBs ($r \gg a_{\rm bin}$), $F(t)$ in eqn.~(\ref{ddho}) is
dominated by GWs. Close to the MBHB, tidal forcing at frequency
$\nu_{\rm GW}=2\nu_{\rm bin}$ is added to $F(t)$ via the Newtonian
quadrupole potential ($F_{\rm NQ} \sim G\mu_{\rm bin}a^{2}_{\rm bin}
R_{\ast}D_{\ast}^{-5}$) and relativistic current dipole force ($F_{\rm
  CD} \sim (G/c)\mu_{\rm bin} v_{\rm bin} a_{\rm
  bin}R_{\ast}D_{\ast}^{-4}$) \citep{MTW}(see eqn. (2.15) in
\citep{Alvi00} and eqn. (6.4) in \citep{Johnson09}). Compared to GW
forcing,
\begin{equation}
\frac{F_{\rm NQ}}{F_{\rm GW}} \approx \left(\frac{\nu_{\rm GW}}{0.02
  \rm{mHz}}\right)^{-4} \left(
\frac{D_{\ast}}{10^{3}r_{g}}\right)^{-4}
\label{nq}
\end{equation}
and
\begin{equation}
\frac{F_{\rm CD}}{F_{\rm GW}} \approx \left( \frac{\nu_{\rm GW}}{0.02
  \rm{mHz}}\right)^{-3} \left(
\frac{D_{\ast}}{10^{3}r_{g}}\right)^{-3}
\label{cd}
\end{equation}
Tidal forcing on a star $10^{3} D_{*,3} r_{g}$ from an MBHB is
dominated by near-field effects at $\lesssim 0.02 D_{*,3}^{-1}$
mHz. The heating from eqn.~(\ref{ssheat}) scales in the near field as
$|F_{\rm NQ}|^{2}/|F_{\rm GW}|^{2}$ and $|F_{\rm CD}|^{2}/|F_{\rm
  GW}|^{2}$.

\section{Electromagnetic observables:} 
GW heating of stars with a large
radiative core causes modest structural changes and increase in
luminosity, since $E_{m}\ll E_{\ast}$. For fully (or mostly)
convective stars (e.g. M-stars), $E_{m}$ is transfered to high-degree
modes concentrated in the outer convective skin, with small mass
$M_{\rm out}\ll M_\ast$ \citep{kumargoodman96}. If $E_{m}>(M_{\rm
  out}/M_{\ast})E_\ast$, the binding energy of the surface skin, the
skin can expand \citep{pod96}, provided $E_{m}$ is thermalized faster
than $t_{\rm conv}$, i.e. for $E_{m}\gsim 10^{45}$erg.  If
$10^{-3}M_{\odot}$ is shed from a Sun-like star and subsequently
accreted onto a $10^6{\rm M_\odot}$ MBHB over a $\sim$year (or
$\approx$10 stellar orbits at $10^3r_g$), the MBHB is fuelled at
0.01-0.1 its Eddington rate (for 10\% radiative efficiency)
\citep{hay13,dai13}.  During accretion, the
MBHB period appears as see-saw variability in the wings of broad
emission lines \citep{mck13}, possibly preceeding tidal disruption
events by MBHBs.

If $E_m\lsim 10^{45}$ erg is thermalized, the star may not bloat but
the luminosity $L^{\prime}_{\ast} =E_m/\tau_d$ can be large.  For the
$M_1=M_2=10^6~{\rm M_\odot}$ MBHB in Fig.~1, at $\nu\approx 0.3$ mHz,
a star at $D_{\ast}=10^3\,r_g$ is heated for $t_F\approx6$hr, and
$E_m=2\times10^{43}$erg emerges over the nonlinear dissipation
timescale $\tau_d\approx 4$yr. During this period $L^{\prime}_{\ast}
\approx 45 (M_{m}/M_{\ast})^{2}L_\odot$. Moreover, $\dot{\omega}_{\rm
  GW}$ sweeps through a large number of resonant modes $N_{\rm m}$
between $40 \mu{\rm Hz}$--$12\,{\rm mHz}$ in the final 4\,yr before
merger \citep{astseism10}. If $N_{\rm m} \sim 10$ modes can be driven resonantly within a
dissipation timescale, then $L^{\prime}_{\ast} \sim \rm{few}\times
10^{2-3}(N_{\rm m}/10)(M_{m}/M_{\ast})^{2}{\rm L_\odot}$. 

\section{Conclusions.}
\label{sec:conclusions}
Quadrupolar oscillation modes in stars can resonate with incident GWs,
reaching non-linear amplitudes at the expense of GW energy.  The
opacity to GWs is distance-independent, so the Sun can eclipse GW
sources (e.g. WDBs) in the ecliptic plane, imprinting absorption lines
in GW spectra. Stars near MBHBs act as GW-charged batteries, discharging
via a brief, significant luminosity increase in convective stars. Mass
loss from the outer skin of stars yields bursts of near-Eddington
accretion onto nearby MBHB. Detailed numerical studies (including
models of stellar structure, effects of rotation) are needed for more
quantitative predictions.

{\section{Acknowledgements.}} We thank Scott Hughes, Pawan Kumar,
Jeremy Goodman, Cole Miller \& Bernd Schutz for very useful
discussions. BM \& KESF are supported by NASA APRA08-0117, NSF PAARE
AST-1153335 and NSF PHY11-25915. BM acknowledges support from a BMCC Faculty Development Grant and a 
CUNY Chancellor's Research Fellowship. ZH is supported by NASA grant
NNX11AE05G.  BK is supported by the W.M. Keck Foundation Fund of the
Institute for Advanced Study and NASA grant NNX11AF29G


\begin{thebibliography}{99}
\bibitem[\protect\citeauthoryear{Aerts et al.}{2010}]{astseism10} Aerts C., Christensen-Dalsgaard J.\& Kurtz D.W., 2010, Astroseismology, Astronomy and Astrophysics Library, ISBM 978-1-4020-5178-4. Springer. 
\bibitem[\protect\citeauthoryear{Alvi}{2000}]{Alvi00} Alvi K., 2000, PRD, 61, 124013
\bibitem[\protect\citeauthoryear{Amaro-Seoane et al.}{2007}]{Amaro07} Amaro-Seoane P., Gair J.R., Freitag M., Miller M.C., Mandel I., Cutler C. \& Babak S., 2007, CQG, 24, 113
\bibitem[\protect\citeauthoryear{Beky \& Kocsis}{2013}]{bk13} Beky, B. \& Kocsis B., 2013, ApJ, 762, 35
\bibitem[\protect\citeauthoryear{Chandrasekhar \& Ferrari}{1991}]{Chandra91} Chandrasekhar S. \& Ferrari V., 1991, RSPSA, 432, 247
\bibitem[\protect\citeauthoryear{Chandrasekhar \& Ferrari}{1992}]{Chandra92} Chandrasekhar S. \& Ferrari V., 1992, RSPSA, 437, 133
\bibitem[\protect\citeauthoryear{Comerford et al.}{2013}]{Comerford13} Comerford J.M. Schluns K., Greene J. E. \& Cool R.J., 2013, ApJ, 777, 64
\bibitem[\protect\citeauthoryear{Comerford \& Greene}{2014}]{Comerford14} Comerford J.M. \& Greene, 2014, ApJ, 789, 112
\bibitem[\protect\citeauthoryear{Crowder \& Cornish}{2007}]{Crowder07} Crowder J. \& Cornish N.J., 2007, PRD, 75, 043008
\bibitem[\protect\citeauthoryear{Cutler \& Lindblom}{1996}]{CutLind96} Cutler \& Lindblom 1996, PRD, 54, 1287
\bibitem[\protect\citeauthoryear{Cutler}{1998}]{cutler98} Cutler C., 1998, PRD, 57, 7089
\bibitem[\protect\citeauthoryear{Dai, Escala \& Coppi}{2013}]{dai13} Dai L., Escala A. \& Coppi P., 2013, ApJ, 775, L9
\bibitem[\protect\citeauthoryear{Fabbiano et al.}{2011}]{Fab11} Fabbiano G., Wang J., Elvis M. \& Risaliti G., 2011, 
Nature, 477, 431
\bibitem[\protect\citeauthoryear{Fabian \& Gough}{1984}]{FabGough84} Fabian A.C. \& Gough D.O., 1984, 
Nature, 308, 160
\bibitem[\protect\citeauthoryear{Hawking}{1966}]{Hawk66} Hawking S.W., 1966, ApJ, 145, 544
\bibitem[\protect\citeauthoryear{Hayasaki, Stone \& Loeb}{2013}]{hay13} Hayasaki K., Stone N. \& Loeb A., 2013, MNRAS, 434, 909
\bibitem[\protect\citeauthoryear{Johnson-McDaniel et al.}{2009}]{Johnson09} Johnson-McDaniel N.K., Yunes N., Tichy W. \& Owen B.J., 2009, PRD, 80, 124039
\bibitem[\protect\citeauthoryear{Khosroshahi \& Sobouti}{1997}]{KS97} Khoshroshahi H.G. \& Sobouti Y., 1997, A\&A, 321, 1024
\bibitem[\protect\citeauthoryear{Kocsis \& Loeb}{2008}]{KocsisLoeb08} Kocsis B. \& Loeb A., 2008, PRL, 101, 1101
\bibitem[\protect\citeauthoryear{Kojima \& Tanimoto}{2005}]{Kojima05} Kojima Y. \& Tanimoto H., 2005, 
CQGrav, 22, 2991
\bibitem[\protect\citeauthoryear{Komossa et al.}{2013}]{Kom03} Komossa S., Burwitz V., Hasinger G., Predehl P., Kaastra J.S. \& Ikebe Y., 2003, ApJ, 582, L15
\bibitem[\protect\citeauthoryear{Kormendy \& Ho}{2013}]{KormHo13} Kormendy J. \& Ho L.C., 2013, ARA\&A, 51, 511
\bibitem[\protect\citeauthoryear{Kumar \& Goodman}{1996}]{kumargoodman96} Kumar P. \& Goodman J., 1996, 466, 946
\bibitem[\protect\citeauthoryear{Li, Kocsis \& Loeb}{2012}]{Li12} Li G., Kocsis B. \& Loeb A., 2012, MNRAS, 425, 2407
\bibitem[\protect\citeauthoryear{Lopes \& Silk}{2014}]{LopesSilk14} Lopes I. \& Silk J., 2014, ApJ (submitted), arXiv:1405.0292 
\bibitem[\protect\citeauthoryear{McKernan \& Yaqoob}{1998}]{MY98} McKernan \& Yaqoob 1998, ApJ, 501, L29
\bibitem[\protect\citeauthoryear{McKernan et al. 2012}{2012}]{McK12} McKernan B., Ford K.E.S., Lyra W. \& Perets H.B., 2012, MNRAS, 425, 460
\bibitem[\protect\citeauthoryear{McKernan et al. 2014}{2014}]{McK14} McKernan B., Ford K.E.S., Kocsis B., Lyra W. \& Winter L.M., 2014, MNRAS, 441, 900
\bibitem[\protect\citeauthoryear{McKernan et al.}{2013}]{mck13} McKernan B., Ford K.E.S., Kocsis B. \& Haiman Z., 2013, MNRAS, 432, 1468
\bibitem[\protect\citeauthoryear{Misner, Thorne \& Wheeler}{1973}]{MTW} Misner C. W., Thorne K.S. \& Wheeler J.A., 1973, Gravitation, Freeman \& Co. 
\bibitem[\protect\citeauthoryear{Peters \& Mathews}{1963}]{Peters&Mathews63} Peters P.C. \& Mathews J., 1963, Physical Review, 131, 435 
\bibitem[\protect\citeauthoryear{Podsiadlowski}{1996}]{pod96} Podsiadlowski P., 1996, MNRAS, 279, 1104
\bibitem[\protect\citeauthoryear{Press \& Teukolsky}{1977}]{pt77} Press W.H. \& Tuekolsky S.A., 1977, ApJ, 213, 183
\bibitem[\protect\citeauthoryear{Rathore et al.}{2005}]{Rathore05} Rathore Y., Blandford R.D. \& Broderick A. E., 2005, MNRAS, 357, 834
\bibitem[\protect\citeauthoryear{Ray, Kembhavi \& Antia}{1987}]{Rayetal1987} Ray A., Kembhavi A.K. \& Antia H.M., 1987, A\&A, 184, 164 
\bibitem[\protect\citeauthoryear{Reisenegger \& Goldreich}{1994}]{rg94} Reisenegger A. \& Goldreich P., 1994, ApJ, 426, 688
\bibitem[\protect\citeauthoryear{Rodriguez et al.}{2006}]{Rod06} Rodriguez C., Taylor G.B., Zavala R.T., Peck A.B., Pollack L.K. \& Romani R.W., 2006, ApJ, 646, 49
\bibitem[\protect\citeauthoryear{Siegel \& Roth}{2010}]{siegelroth10} Siegel D.M. \& Roth M., 2010, MNRAS, 408, 1742
\bibitem[\protect\citeauthoryear{Siegel \& Roth}{2011}]{siegelroth11} Siegel D.M. \& Roth M., 2011, ApJ, 729, 137
\bibitem[\protect\citeauthoryear{Stix et al.}{1993}]{Stix+1993} Stix M., Rudiger G., Knolker M. \& Grabowski U., 1993, A\&A, 272, 340
\bibitem[\protect\citeauthoryear{Timpano et al.}{2006}]{Timp06} Timpano S.E., Rubbo L.J. \& Cornish N.J., 2006, PRD, 73, 122001
\bibitem[\protect\citeauthoryear{Thorne \& Braginsky}{1976}]{ThBrag76} Thorne K.S. \& Braginsky V.B., 1976, ApJ, 204, L1
\bibitem[\protect\citeauthoryear{Turner \& Miller}{2009}]{TM09} Turner \& Miller 2009, A\&A Rv, 17, 47.
\bibitem[\protect\citeauthoryear{Weinberg et al.}{2012}]{W12} Weinberg N., Arras P., Quataert E. \& Burkart J., 2012, ApJ, 751, 136
\bibitem[\protect\citeauthoryear{Woo et al.}{2014}]{Woo14} Woo J.-H., Cho H., Husemann B., Komossa S., Park D. \& Bennert V.N., 2014, MNRAS, 437, 32
\end{thebibliography}
\end{document}